\documentclass[aps,showplace,twocolumn,preprintnumbers,bibnotes]{revtex4}
\usepackage{ifpdf}
\usepackage{graphicx}
\usepackage{color}
\usepackage{dcolumn}
\usepackage{bm}
\usepackage{mathrsfs}
\usepackage{amsfonts}
\usepackage{amssymb}
\usepackage{amsmath}
\usepackage{dsfont}
\usepackage{longtable}
\usepackage{fixmath}
\usepackage{upgreek}
\usepackage{url}
\usepackage[bookmarks,colorlinks]{hyperref}
\hypersetup{colorlinks=true,%
            citecolor=blue,%
            filecolor=black,%
            linkcolor=blue,%
            urlcolor=blue}
\def\t#1{\tilde{#1}}
\def\wh#1{\widehat{#1}}
\def\h#1{\hat{#1}}

\def\Sc#1{\textsc{#1}}

\DeclareMathOperator{\re}{Re}
\DeclareMathOperator{\im}{Im}
\DeclareMathOperator{\e}{e}
\DeclareMathOperator{\rd}{d\!}

\begin{document}
\ifx\href\undefined\else\hypersetup{linktocpage=true}\fi
\bibliographystyle{apsrev}

\preprint{Preprint number: ITP-UU-2008/24}

\title{Comment on ``Violation of the Luttinger sum rule within the Hubbard model on a triangular lattice'', by J. Kokalj and P. Prelov\v{s}ek, \href{http://dx.doi.org/10.1140/epjb/e2008-00264-0}{\emph{Eur. Phys. J.} B~{\bf 63}, 431 (2008)} }

\author{Behnam Farid}
\affiliation{Institute for Theoretical Physics, Department of Physics and Astronomy, University of Utrecht, \\
Leuvenlaan 4, 3584 CE Utrecht, The Netherlands }
\email{B.Farid@phys.uu.nl}

\date{\today}

\begin{abstract}
Using the first-order series expansion of the function $G({\bm k};\mu)$, in powers of $\t{\mu} \doteq \mu-U/2$, pertaining to the insulating ground state of a single-band Hubbard Hamiltonian at half-filling, Kokalj and Prelov\v{s}ek (KP) have in a recent paper [\emph{Eur. Phys. J.} B~{\bf 63}, 431 (2008)] reported breakdown of the Luttinger theorem for the specific case where the lattice on which the Hubbard Hamiltonian is defined is a two-dimensional \emph{triangular} lattice, for which the ground state is not invariant under particle-hole transformation. Here $G({\bm k};\mu)$ is the single-particle Green function $G({\bm k};\varepsilon)$ evaluated at $\varepsilon=\mu$, the zero-temperature limit of the chemical potential corresponding to half-filling, and $U$ the on-site interaction energy. In this Comment we demonstrate that unless $\t{\mu} = 0$ (to be strictly distinguished from $\t{\mu}$ small but non-vanishing), any \emph{finite-order} series expansion for $G({\bm k};\mu)$ in powers of $\t{\mu}$ in general falsely signals breakdown of the Luttinger theorem. The violation of this theorem as asserted by KP is therefore an artifact of their first-order calculation.
\end{abstract}


\maketitle



Considering the non-magnetic uniform ground state (GS) of Hamiltonian $\wh{H}$, corresponding to $N$ fermions, the associated single-particle Green function $G^{\sigma}({\bm k};\varepsilon)$ is independent of the spin index $\sigma$. Consequently, in this Comment we suppress $\sigma$ and denote the latter function by $G({\bm k};\varepsilon)$. For the mentioned $N$-particle GS, the Luttinger theorem under consideration states that \cite{LW60,JML60,IED03,AMT03,ET05,KRT06,BF07a,BF07b}
\begin{equation}
N = 2 \sum_{\bm k} \Theta\big(G({\bm k};\mu)\big), \label{e1}
\end{equation}
where $\mu$ is the zero-temperature limit of the chemical potential satisfying the equation of state of the grand-canonical ensemble in which the mean number of particles is equal to $N$ \cite{BF07a}. The summation with respect to ${\bm k}$ in Eq.~(\ref{e1}) is over the entire wave-vector space available to the system under consideration. For instance, for $\wh{H}$ defined on a Bravais lattice, $\{ {\bm R}_i\}$, this space consists of the points constituting the corresponding first Brillouin zone, $\mathrm{1BZ}$.

In the thermodynamic limit, where the underlying ${\bm k}$ space consists of a continuum, one defines the Luttinger surface $\mathcal{S}_{\Sc l}$ corresponding to the insulating GS of $\wh{H}$ as the locus of the ${\bm k}$ points for which
\begin{equation}
G({\bm k};\mu) = 0. \label{e2}
\end{equation}
We point out that Eq.~(\ref{e2}) should be used with caution, as $\mathcal{S}_{\Sc l}$ has a more encompassing definition than implied by this equation (Sec.~2.4 in Ref.~\cite{BF07a}). For the considerations in this Comment, we shall not strive generality and therefore in the following consider Eq.~(\ref{e2}) as the defining equation for $\mathcal{S}_{\Sc l}$.

The Luttinger surface separates the ${\bm k}$ space into a region whose points contribute to the Luttinger sum on the right-hand (RHS) of Eq.~(\ref{e1}), i.e. the Luttinger sea, and a complementary region. For the two-dimensional system that we consider here, it turns out that the underlying $\mathcal{S}_{\Sc l}$ consists of one piece, and so does the Luttinger sea.

For the economy of notation, we introduce the radial unit vector $\h{\bm n}$ centred at the origin of the ${\bm k}$ space (here the $\mathrm{1BZ}$) and directed along the radial vector ${\bm k}$ under consideration, whereby we can write
\begin{equation}
{\bm k} = k\, \h{\bm n}, \label{e3}
\end{equation}
where $k\ge 0$. We thus define
\begin{equation}
G^{(m,n)}({\bm k};\mu) \doteq \frac{\partial^m}{\partial k^m} \frac{\partial^n}{\partial \mu^n} G({\bm k};\mu). \label{e4}
\end{equation}

With $\mu$ denoting the exact zero-temperature limit of the chemical potential corresponding to $N$ particles, for an insulating GS KP \cite{KP07} employed the first-order approximation of the exact Taylor expansion
\begin{equation}
G({\bm k};\mu) = \sum_{n=0}^{\infty} \frac{G^{(0,n)}({\bm k};U/2)}{n!}\, \t{\mu}^n, \label{e5}
\end{equation}
where
\begin{equation}
\t{\mu} \doteq \mu -\frac{1}{2} U. \label{e6}
\end{equation}
The series in Eq.~(\ref{e5}) being a power series, its validity is restricted to the region where $\vert\t{\mu}\vert$ is less that the smallest of the two positive numbers $\varepsilon_{+} - U/2$ and $U/2 -\varepsilon_{-}$, in which $\varepsilon_{+}$ ($\varepsilon_{-}$) is the smallest (largest) energy $\varepsilon$ greater (less) than $\mu$ for which $\sum_{\bm k} A({\bm k};\varepsilon)$ is non-vanishing; here
\begin{equation}
A({\bm k};\varepsilon) \doteq \mp \frac{1}{\pi} \im[G({\bm k};\varepsilon\pm i 0^+)] \label{e7}
\end{equation}
is the single-particle spectral function.

More explicitly, by employing the spectral representation for $G({\bm k};\mu)$ in terms of $A({\bm k};\varepsilon)$, KP \cite{KP07} expressed $G({\bm k};\mu)$ as
\begin{equation}
G({\bm k};\mu) \equiv G^{-}({\bm k};\mu) + G^{+}({\bm k};\mu), \label{e8}
\end{equation}
in which $G^{\mp}({\bm k};\mu)$ arise from the spectral contributions corresponding to $\varepsilon \lessgtr \mu$. Using the standard moments expansion of the underlying integrals \cite{ST70,HL67}, KP obtained that \cite{KP07}
\begin{equation}
G^{\mp}({\bm k};\mu) = \sum_{n=0}^{\infty} \Big(\!\!\pm\frac{2}{U}\Big)^{n+1} \sum_{m=0}^n M^{\mp}_{n-m}({\bm k}) \binom{n}{m} (-\t{\mu})^m, \label{e9}
\end{equation}
where the explicit expressions for $M_{l}^{\mp}({\bm k})$ are given in Eq.~(6) and (7) of Ref.~\cite{KP07}.

Making use of the identity
\begin{equation}
\sum_{n=0}^{\infty} \sum_{m=0}^{n} f_{n,m} \equiv \sum_{n=0}^{\infty} \sum_{m=n}^{\infty} f_{m,n}, \label{e10}
\end{equation}
the expressions in Eq.~(\ref{e9}) can be equivalently expressed as
\begin{equation}
G^{\mp}({\bm k};\mu) =\! \sum_{n=0}^{\infty} \!\Big[ (-1)^{n} \!\sum_{m=n}^{\infty} \!\Big(\pm\frac{2}{U}\Big)^{m+1}\! M^{\mp}_{m-n}({\bm k}) \binom{m}{n}\Big]\, \t{\mu}^{n}, \label{e11}
\end{equation}
from which the explicit expression for $G^{(0,n)}({\bm k};U/2)$ can be immediately read off. We point out that since KP \cite{KP07} considered the problem at hand in the regime $\vert t/U\vert \ll 1$, in their explicit calculations they employed expressions for $G^{(0,n)}({\bm k};U/2)$, $n=0, 1$, that are correct to the leading non-trivial order in $t/U$. This amounts to a specific truncation of the sum with respect to $m$ on the RHS of Eq.~(\ref{e11}). We shall briefly return to this aspect later.

Denoting by $G_{\nu}({\bm k};\mu)$ the function that one obtains on replacing the upper bound of the sum on the RHS of Eq.~(\ref{e5}) by $\nu$, one has
\begin{eqnarray}
&&\hspace{-1.2cm} G({\bm k};\mu) \equiv G_{\nu}({\bm k};\mu) + \sum_{n=\nu+1}^{\infty}  \frac{G^{(0,n)}({\bm k};U/2)}{n!}\, \t{\mu}^n \nonumber\\
&&\hspace{-1.0cm} \sim G_{\nu}({\bm k};\mu) +  \frac{G^{(0,\nu+1)}({\bm k};U/2)}{(\nu+1)!}\, \t{\mu}^{\nu+1} \;\;\, \mbox{\rm for}\;\;\, \t{\mu} \to 0. \label{e12}
\end{eqnarray}

Since $G({\bm k};\mu)$ is unknown, within the approximate framework adopted by KP \cite{KP07} one calculates an approximate Luttinger surface, $\mathcal{S}_{\Sc l}^{(\nu)}$, through solving the equation
\begin{equation}
G_{\nu}({\bm k};\mu) = 0, \label{e13}
\end{equation}
where $\nu = 1$ for the explicit calculations reported by KP \cite{KP07}. For a predetermined $\h{\bm n}$, we denote the solution of Eq.~(\ref{e13}) by ${\bm k}_{\Sc l}^{(\nu)}$, and, with reference to the convention in Eq.~(\ref{e3}), the corresponding $k$ by $k_{\Sc l}^{(\nu)}$. Similarly, for the solution of Eq.~(\ref{e2}) in the direction of the same $\h{\bm n}$, which we denote by ${\bm k}_{\Sc l}$, and the corresponding $k$ by $k_{\Sc l}$. Following Eqs.~(\ref{e12}) and (\ref{e13}) one thus has
\begin{equation}
G({\bm k}_{\Sc l}^{(\nu)};\mu) \sim \frac{G^{(0,\nu+1)}({\bm k}_{\Sc l}^{(\nu)};U/2)}{(\nu+1)!}\, \t{\mu}^{\nu+1} \;\;\, \mbox{\rm for}\;\;\, \t{\mu} \to 0. \label{e14}
\end{equation}

Since ${\bm k}_{\Sc l}^{(\nu)} \to {\bm k}_{\Sc l}$ as $\t{\mu} \to 0$, by assuming that $G({\bm k};\mu)$ is regular for ${\bm k}$ in a finite neighbourhood of $\mathcal{S}_{\Sc l}$, one can expand $G({\bm k}_{\Sc l}^{(\nu)};\mu)$ in powers of $k_{\Sc l}^{(\nu)} - k_{\Sc l}$ for sufficiently small values of $\t{\mu}$. Since by definition $G({\bm k}_{\Sc l};\mu) = 0$, Eq.~(\ref{e2}), one thus obtains that
\begin{equation}
G({\bm k}_{\Sc l}^{(\nu)};\mu) = \sum_{m=1}^{\infty} \frac{G^{(m,0)}({\bm k}_{\Sc l};\mu)}{m!}\, (k_{\Sc l}^{(\nu)} - k_{\Sc l})^m. \label{e15}
\end{equation}
Truncating this series at $m=1$, and assuming that $G^{(1,0)}({\bm k}_{\Sc l};\mu) \not=0$, from the resulting expression and the asymptotic result in Eq.~(\ref{e14}) one deduces that
\begin{equation}
k_{\Sc l}^{(\nu)} \sim k_{\Sc l} +\frac{G^{(0,\nu+1)}({\bm k}_{\Sc l}^{(\nu)};U/2)}{G^{(1,0)}({\bm k}_{\Sc l};\mu)} \frac{\t{\mu}^{\nu+1}}{(\nu+1)!}\;\;\, \mbox{\rm for}\;\;\, \t{\mu} \to 0. \label{e16}
\end{equation}
For sufficiently small values of $\t{\mu}$, one can replace the ${\bm k}_{\Sc l}$ on the RHS of this expression by ${\bm k}_{\Sc l}^{(\nu)}$, and \emph{vice versa}.

The simple asymptotic expression in Eq.~(\ref{e16}) makes explicit that unless $\t{\mu} = 0$, $k_{\Sc l}^{(\nu)}$ does \emph{not} coincide with $k_{\Sc l}$ for any \emph{finite} value of $\nu$. Evidently, it is in principle possible that $k_{\Sc l}^{(\nu)}$ may be equal to $k_{\Sc l}$ for \emph{some} directions of $\h{\bm n}$, however this equality cannot apply for all directions of $\h{\bm n}$. Although in spite of $k_{\Sc l}^{(\nu)} \not= k_{\Sc l}$, the Luttinger theorem, Eq.~(\ref{e1}), may apply when the $G({\bm k};\mu)$ on the RHS of Eq.~(\ref{e1}) is replaced by $G_{\nu}({\bm k};\mu)$, with $\nu<\infty$ (see Fig.~47, p.~231, in Ref.~\cite{PN64}), this need not be the case in general. For instance, in the cases where both $\mathcal{S}_{\Sc l}$ and $\mathcal{S}_{\Sc l}^{(\nu)}$, with $\nu <\infty$, are isotropic or nearly isotropic, the Luttinger theorem in terms of $G_{\nu}({\bm k};\mu)$ unquestionably fails. We have thus unequivocally demonstrated that \emph{for $\nu<\infty$, the strategy adopted by KP in Ref.~\cite{KP07} is not appropriate for verifying the validity of the Luttinger theorem}.

We shall now focus on the numerical results reported by KP \cite{KP07}. As we have indicated earlier, these results correspond to $\nu=1$. For the insulating GS of a single-band Hubbard Hamiltonian, in Ref.~\cite{KP07} (Eq.~(16) herein) KP obtained that
\begin{equation}
G({\bm k};\mu) = \frac{4}{U^2} \big(\sum_{\delta} 4 C_{\delta}\, \varepsilon_{\delta}({\bm k}) - \t{\mu}\big) + O\big(\frac{t_{i,j}^2}{U^3}\big), \label{e17}
\end{equation}
where \cite{KP07}
\begin{equation}
C_{\delta} \doteq \langle {\bm S}_{\delta} \cdot {\bm S}_0\rangle
\label{e18}
\end{equation}
is the GS spin-spin correlation function, and
\begin{equation}
\varepsilon_{\delta}({\bm k}) \doteq - \sum_{i_{\delta}} t_{i_{\delta},0}\, \e^{i {\bm k}\cdot {\bm R}_{i_{\delta}}}, \label{e19}
\end{equation}
in which $\delta \in \{ 1,2, \dots\}$ refers to nearest-neighbour, next-nearest-neighbour, \dots, so that  $\sum_{i_{\delta}}$ is a sum over the lattice vectors $\{{\bm R}_{i_{\delta}} \}$ which are nearest-neighbours, next-nearest-neighbours, \dots, to the central lattice vector ${\bm 0}$ for $\delta = 1, 2, \dots$, respectively. With reference to the remark following Eq.~(\ref{e11}) above, we note that the error $O(t_{i,j}^2/U^3)$ on the RHS of Eq.~(\ref{e17}) corresponds to restricting the upper bound of the sum with respect to $m$ on the RHS of Eq.~(\ref{e11}) to $1$.

Specialising to the cases where the hopping integral $t_{i,j}$ is non-vanishing only for nearest-neighbour sites (assuming further that $t_{i,j} = t$ for ${\bm R}_i$ and ${\bm R}_j$ nearest neighbours) and suppressing the last term on the RHS of Eq.~(\ref{e17}), KP \cite{KP07} employed the following simple expression for $U/t=40$ in examining the validity of the Luttinger theorem \cite{Note1}:
\begin{equation}
G({\bm k};\mu) = \frac{4}{U^2} \big(4 C_1\, \varepsilon_1({\bm k}) - \t{\mu}\big) \equiv \frac{16 C_1}{U^2} \big( \varepsilon_1({\bm k}) -\frac{\t{\mu}}{4 C_1}\big), \label{e20}
\end{equation}
where \cite{KP07}
\begin{equation}
C_1 = -0.182, \label{e21}
\end{equation}
and (identifying the lattice constant $a$ with unity)
\begin{equation}
\varepsilon_1({\bm k}) = -2 t \big(\! \cos(k_x) + 2 \cos(\frac{1}{2}\, k_x) \cos(\frac{\sqrt{3}}{2}\, k_y) \big), \label{e22}
\end{equation}
in which $k_x$ and $k_y$ are the Cartesian coordinates of ${\bm k}$. With reference to Eq.~(\ref{e2}), one observes that according to Eq.~(\ref{e20}) for the system under consideration the $\mathcal{S}_{\Sc l}$ (i.e. the $\mathcal{S}_{\Sc l}^{(1)}$) corresponding to $\vert U/t\vert \gg 1$ coincides with the Fermi surface $\mathcal{S}_{\Sc f}^0$ of the non-interacting system, associated with $\varepsilon_1({\bm k})$ and corresponding to the chemical potential $\mu_0$, where
\begin{equation}
\mu_0 \doteq \frac{\t{\mu}}{4 C_1}. \label{e23}
\end{equation}
The value of $\mu_0$ corresponding to half-filling is obtained by solving the equation
\begin{equation}
\frac{1}{\mathscr{S}_{1\Sc b\Sc z}} \int_{\Sc 1\Sc b\Sc z} {\rm d}^2k\; \Theta\big(\mu_0 - \varepsilon_1({\bm k})\big) = \frac{1}{2}, \label{e24}
\end{equation}
where $\mathscr{S}_{1\Sc b\Sc z} = 8\pi^2/\sqrt{3}$ is the area of the 1BZ under consideration in the units where the lattice constant $a$ is identified with unity; this 1BZ is a regular hexagon \cite{CL00,Note2} of which the Cartesian coordinates of the vertices are:
\begin{equation}
2\pi (\frac{2}{3}, 0),\; 2\pi (\frac{1}{3}, \pm\frac{\sqrt{3}}{3}),\; 2\pi (-\frac{1}{3}, \pm\frac{\sqrt{3}}{3}),\; 2\pi (-\frac{2}{3}, 0). \label{e25}
\end{equation}
Numerical calculation yields $\mu_0 \approx 0.852 t$, which following Eq.~(\ref{e23}) leads to \cite{KP07}
\begin{equation}
\t{\mu} = \t{\mu}_1 \doteq 4 C_1 \mu_0 \approx -0.620\, t \label{e26}
\end{equation}
as the value for $\t{\mu}$ required for the Green function $G({\bm k};\mu)$ in Eq.~(\ref{e20}) (i.e. $G_{\nu=1}({\bm k};\mu)$) to satisfy the Luttinger theorem, Eq.~(\ref{e1}).

Through a combination of explicit numerical calculations (see Fig.~1 in Ref.~\cite{KP07}), KP \cite{KP07} deduced the following expression for $\t{\mu}$ specific to half-filled uniform GSs:
\begin{equation}
\t{\mu} \sim - (0.19 \pm 0.1) t + \frac{6.8 t^2}{U}\;\;\; \mbox{\rm for}\;\;\;
0.025 t < \frac{t^2}{U} < 0.05 t. \label{e27}
\end{equation}
With $U/t = 40$, this expression yields
\begin{equation}
\t{\mu} = (-0.02 \pm 0.1)\, t, \label{e28}
\end{equation}
a range of values for $\t{\mu}$ which is far too removed from the value presented in Eq.~(\ref{e26}), i.e. the value required in order for the Luttinger theorem to apply for large values of $\vert U/t\vert$, and sufficiently large systems (for which the substitution of $\sum_{\bm k}$ by $[\mathscr{S}/(2\pi)^2] \int_{1\Sc b\Sc z} {\rm d}^2 k$, where $\mathscr{S}$ is the area of the system, is justified). \emph{Thus, KP \cite{KP07} arrived at the conclusion that for the GS under consideration, the Luttinger theorem broke down.}

The fallacy in the reasoning by KP \cite{KP07} becomes apparent by realising that the Green function in Eq.~(\ref{e20}) concerns $G_{\nu=1}({\bm k};\mu)$ (see Eq.~(\ref{e12})), to be distinguished from the exact $G({\bm k};\mu)$, for large values of $\vert U/t\vert$. With reference to Eq.~(\ref{e16}), it is evident that unless the coefficient function multiplying $\t{\mu}^{\nu+1}/(\nu+1)!$ is vanishing (here $\nu=1$), any non-vanishing value of $\t{\mu}$ implies deviation of the $\mathcal{S}_{\Sc l}^{(\nu)}$ calculated on the basis of $G_{\nu}({\bm k};\mu)$ from the exact $\mathcal{S}_{\Sc l}$. Since in the case at hand $\mathcal{S}_{\Sc l}^{(1)}$ is nearly isotropic (see Fig.~2 in Ref.~\cite{KP07} --- one expects the same to apply for the exact $\mathcal{S}_{\Sc l}$), this deviation directly leads to an apparent, but \emph{false}, violation of the Luttinger theorem.

On general grounds, one can demonstrate that $G^{(0,\nu+1)}({\bm k}_{\Sc l}^{(\nu)};U/2)$ is for insulating GSs relatively large, leading to a considerable deviation of $k_{\Sc l}^{(\nu)}$ from $k_{\Sc l}$ even for relatively small but non-vanishing values of $\vert\t{\mu}\vert$ (see Eq.~(\ref{e16})). To appreciate this fact, one should realise that by the Kramers-Kr\"onig relation for $\re[G(k;\varepsilon)]$, the sudden change of $\im[G(k;\varepsilon\pm i 0^+)]$ for $\varepsilon$ at band edges (from identically vanishing for $\varepsilon$ inside the underlying gap, to a function whose magnitude is steeply increasing for $\varepsilon$ immediately past the band edges) implies very rapid change in $\re[G({\bm k};\varepsilon)]$ for $\varepsilon$ inside the gap region, leading to very large values of $G^{(0,\nu+1)}({\bm k}_{\Sc l}^{(\nu)};U/2)$ \cite{Note3}. Below we specify the scale according to which the adjective `large' is to be understood here.

For a quantitative analysis, we note that for the coefficient of $\t{\mu}^{\nu+1}/(\nu+1)!$ in Eq.~(\ref{e16}), specific to $\nu=1$, one expects that
\begin{equation}
\frac{G^{(0,2)}({\bm k}_{\Sc l}^{(1)};U/2)}{G^{(1,0)}({\bm k}_{\Sc l};\mu)} = O\Big(\frac{1}{\t{\mu}}\Big)\;\;\; \mbox{\rm for}\;\;\; \t{\mu}\not= 0. \label{e29}
\end{equation}
This result follows from the assumption that $G({\bm k};z)$ is a continuously differentiable function of ${\bm k}$ and $z$ in the neighbourhoods of $\mathcal{S}_{\Sc l}$ and $U/2$ respectively, whereby $k_{\Sc l}^{(1)} - k_{\Sc l}$ must to leading order vary \emph{linearly} with $\t{\mu}$ for small values of $\t{\mu}$. On the basis of this observation, the relationship in Eq.~(\ref{e29}) follows immediately on dividing both sides of Eq.~(\ref{e16}) by $\t{\mu}$ (assuming that $\t{\mu}\not= 0$). With reference to the result in Eq.~(\ref{e27}) (or Eq.~(\ref{e28})), the estimate in Eq.~(\ref{e29}) implies that for the case at hand the left-hand side of this equation is to leading order proportional to $10/t$. The scaling of the latter value with $1/t$, to be contrasted with $1/U$, is most significant.

Following the observation that to leading order $(k_{\Sc l}-k_{\Sc l}^{(1)})/k_{\Sc l}$ is proportional to $\t{\mu}/t$, from the expression in Eq.~(\ref{e28}) one concludes that for the case at hand $(k_{\Sc l}-k_{\Sc l}^{(1)})/k_{\Sc l}$ must be of the order of $10$\%, which conforms with the value of $17$\% as reported by KP \cite{KP07}.

We thus conclude that the inference by KP \cite{KP07}, that the Luttinger theorem broke down for the particle-hole asymmetric insulating GSs of the single-band Hubbard Hamiltonian, is \emph{incorrect}. The reported deviation of the calculated Luttinger surface from the Luttinger surface for which the Luttinger theorem would apply (for large values of $U/t$), is accounted for by the truncation error associated with the use of the first-order expansion for $G({\bm k};\mu)$ in powers of $\t{\mu} \equiv \mu-U/2$. The attempt by KP \cite{KP07} at finding the exact solutions of Eq.~(\ref{e1}) (the totality of which comprise the exact $\mathcal{S}_{\Sc l}$) on the basis of the approximation $G({\bm k};\mu) \approx G_{\nu=1}({\bm k};\mu)$, which applies for a relatively small neighbourhood of $\t{\mu}=0$, may be likened with that of finding the solutions of $\sin(x)/x = 0$ at $x=\pm\pi$ on the basis of the approximation $\sin(x)/x \approx 1 - x^2/6$, which applies for a small neighbourhood of $x=0$; the solutions of $1-x^2/6 = 0$ are $x=\pm\sqrt{6} \approx \pm 2.45$, which are indeed considerably removed from $\pm\pi$ (explicitly, by some $22$\%, which is comparable with the $17$\% reported by KP \cite{KP07}).

In view of the above observations, we wish to close this Comment by a critical message. As the detailed considerations in Ref.~\cite{BF07a} have shown, the Luttinger theorem \emph{is} valid under the conditions specified by Luttinger and Ward \cite{LW60}. In Ref.~\cite{BF07a} considerable amount of space was devoted to demonstrating that a vast body of the extant publications, that supposedly demonstrate failure of the Luttinger theorem, are distinctly \emph{erroneous}. This statement equally applies not only to Ref.~\cite{KP07}, considered here, but also to a most recent publication by KP \cite{KP08} to which a separate Comment \cite{FT1D09}, by the present author and A.~M. Tsvelik, is directed. To our best knowledge, to this date \emph{no} claim of the supposed failure of the Luttinger theorem has withstood the test of time, so that it seems high time that henceforth researchers leave this theorem undisturbed, and resist the temptation of apparently interminably declaring it as invalid through mistaking the imperfections of their pertinent calculations with the failure of this perfectly valid theorem.
\hfill $\square$




\bibliographystyle{apsrev}


\end{document}